\documentclass[conference]{IEEEtran}
\IEEEoverridecommandlockouts
\usepackage{cite}
\usepackage{amsmath,amssymb,amsfonts}
\usepackage{algorithmic}
\usepackage{geometry}
\usepackage{tabularx}
\usepackage{graphicx}
\usepackage{algorithm}
\usepackage{algorithmic}
\usepackage{subfigure}  %插入多图时用子图显示的宏包
\usepackage{float}  %插入多图时用子图显示的宏包
\usepackage{textcomp}
\usepackage{booktabs}
\usepackage{xcolor}
\usepackage{color}
\usepackage{graphicx} %use graph format
\usepackage{epstopdf}
\makeatletter
\renewcommand{\maketag@@@}[1]{\hbox{\m@th\normalsize\normalfont#1}}%
\makeatother
\def\BibTeX{{\rm B\kern-.05em{\sc i\kern-.025em b}\kern-.08em
    T\kern-.1667em\lower.7ex\hbox{E}\kern-.125emX}}

\begin{document}

\title{ISAC 4D Imaging System Based on 5G Downlink Millimeter Wave Signal\\
	\thanks{This work was supported in part by the National
		Key Research and Development Program under Grant 2020YFA0711302,
		in part by the National Natural Science Foundation of China (NSFC) under Grant 62271081, 92267202, and U21B2014.}
}

\author{\IEEEauthorblockN{Bohao Lu$^1$, Zhiqing Wei$^{1,*}$, Lin Wang$^1$, Ruiyun Zhang$^1$, Zhiyong Feng$^1$}
	%\text{Key Laboratory of Universal Wireless Communications, Ministry of Education,}\\
	$^1$\text{Beijing University of Posts and Telecommunications, Beijing 100876, China}\\
	$^1$Email: \{bohaolu, weizhiqing, wlwl, zhangruiyun, fengzy\}@bupt.edu.cn \\}

\maketitle

\begin{abstract}
Integrated Sensing and Communication(ISAC) has become a key technology for the 5th generation (5G) and 6th generation (6G) wireless communications due to its high spectrum utilization efficiency. Utilizing infrastructure such as 5G Base Stations (BS) to realize environmental imaging and reconstruction is important for promoting the construction of smart cities. Current 4D imaging methods utilizing Frequency Modulated Continuous Wave (FMCW) based Fast Fourier Transform (FFT) are not suitable for ISAC scenarios due to the higher bandwidth occupation and lower resolution. We propose a 4D (3D-Coordinates, Velocity) imaging method with higher sensing accuracy based on 2D-FFT with 2D-MUSIC utilizing standard 5G Downlink (DL) millimeter wave (mmWave) signals. To improve the sensing precision we also design a transceiver antenna array element arrangement scheme based on MIMO virtual aperture technique. We further propose a target detection algorithm based on multi-dimensional Constant False Alarm (CFAR) detection, which optimizes the ISAC imaging signal processing flow and reduces the computational pressure of signal processing. Simulation results show that our proposed method has better imaging results. The code is publicly available at \text{https://github.com/MrHaobolu/ISAC\_4D\_IMaging.git}.
\end{abstract}


\section{Introduction}
\subsection{Background and Motivation}
The development of 5G and 6G wireless communications toward higher frequency bands and larger bandwidths has significantly improved the sensing ability of communication signals. ISAC-based technology using widely available wireless communication signals to achieve imaging and 4D reconstruction of urban environment is important for digital twin and smart city construction\cite{wang2023coherent}. Compared with traditional imaging with additional optical detection equipment, ISAC-based wireless imaging has low-deployment-cost, privacy protection, and robustness in severe weather.

\subsection{Related Works}
The majority of current work in wireless imaging focuses on 3D or 4D imaging using Frequency Modulated Continuous Wave (FMCW) radar, which primarily uses FFT-based algorithms for echo signal processing~\cite{sun2020mimo}. Jiang et al.~\cite{jiang20234d}proposed a wireless imaging method that embeds dual pulse repetition frequency (dual-PRF) waveforms into a time-division multiplexing \& Doppler-division multiplexing MIMO (TDM-DDM-MIMO) framework and a super-resolution Direction of Arrival (DoA) estimation Depth Convolutional Network: CV-DCN to obtain high-resolution 4D point cloud images. Santra et al.~\cite{santra2020ambiguity} proposed a MIMO radar imaging scheme based on roughly orthogonal FMCW waveforms. Sun et al.~\cite{sun20214d} proposed a high-resolution imaging radar system using the virtual 2D sparse array. The above FFT-base 4D imaging algorithms have limited resolution compared to Multiple Signal Classification (MUSIC)-based algorithms. Besides, the FMCW is not suitable for ISAC because of its poor communication capability.

Orthogonal frequency division multiplexing (OFDM)-based ISAC waveform is widely studied\cite{chen2023multiple}, but there are few related works on OFDM radar-based 4D imaging algorithms. Guan et al.~\cite{guan20213} built a high-resolution 3D radar imaging system using 5G millimeter wave (mmWave) signals. There is a lack of estimation of targets' velocities and the use of FFT-based signal processing algorithms leads to a significantly limited 3D point cloud resolution among the above work.

In order to improve the resolution of DoA estimation and reduce the false alarm probability of 4D point cloud detection, the virtual aperture technique\cite{stolz2018new}  with Constant False Alarm Rate (CFAR) \cite{gao2019experiments} detection has been introduced in related research. However, the virtual aperture transceiver antenna design for DL active sensing of 5G NR BS is not well designed, and the current widely used CFAR detection algorithms cannot overcome the small target missing detection problem caused by strong interference. Therefore, the improvement of transceiver antenna design and CFAR detection algorithm is also a key issue to achieve 4D imaging in ISAC scenarios.

\subsection{Contributions}
To solve the above problems, we propose an OFDM DL ISAC imaging algorithm based on FFT and MUSIC using 5G mmWave signals. The main contributions and innovations of this paper are summarized as follows.
\begin{itemize}
    \item We propose a BS-side active 4D imaging sensing algorithm using standard 5G NR mmWave signal containing Comb4 structured Positioning Reference Signal (PRS). The proposed algorithm can simultaneously estimate the range, velocity, and azimuth \& pitch angle information (4D information) for a large number of scatter points.
    \item We propose a transceiver antenna array design based on virtual aperture technology, which can effectively improve the resolution of DoA estimation by increasing the virtual array when the size of transceiver antenna array in the current BS is limited.
    \item  We propose a two-stage scattering point detection algorithm based on CFAR detection. The proposed algorithm uses OSCA (Ordered Statistics \& Cell Averaging)-CFAR detection algorithm for the Range-Doppler Map (RDM) obtained by 2D-FFT and CA-CFAR detection algorithm for 2D-MUSIC DoA estimation results, which can reduce the computational complexity and reduce the probability of false alarm compared to the single constant threshold detection method.
\end{itemize}

\section{System And Signal Models}

\subsection{DL ISAC Imaging Model}
\begin{figure}[!htb]
	\centering
	\includegraphics[width=0.5\linewidth]{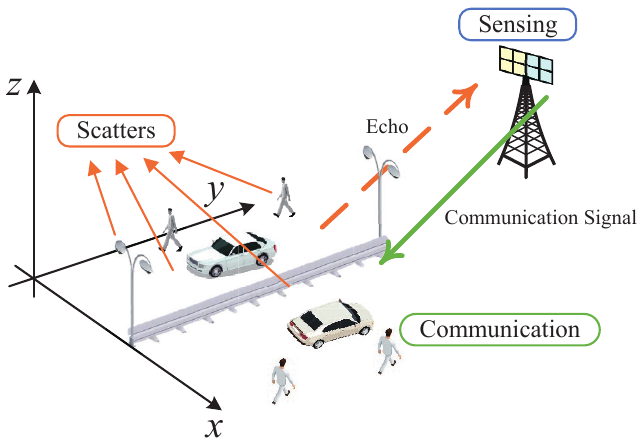}
	\caption{DL ISAC Imaging Scenario.}
	\label{fig:Scenario}
\end{figure}
As shown in Fig.~\ref{fig:Scenario}, we consider the DL ISAC imaging scenario between the 5G BS, the User Equipment (UE), and the user's environment. The scatterers in the environment include vehicles, pedestrians, road infrastructure, etc. The 5G BS is equipped with two spatially well-separated Uniform Planar Arrays (UPAs). At the BS side, one UPA is used to transmit DL ISAC signals and the other UPA is used to continuously receive the echoes of ISAC signals for estimating the range, velocity, and DoAs of the ambient scatterers for 4D imaging. At the UE side, the UE receives the ISAC signals to demodulate the communication data. The transmit and receive arrays of the BS have dimensions ${{P}_{t}}\times {{Q}_{t}}$ and ${{P}_{r}}\times {{Q}_{r}}$, whose array element spacing and the arrangement will be described in detail below.

\subsection{UPAs Model and Virtual Aperture}
\begin{figure}[!htb]
	\setlength {\abovecaptionskip} {0.cm} 
	\setlength {\belowcaptionskip} {-0.cm}
	\centering  %图片全局居中
	\subfigbottomskip=2pt %两行子图之间的行间距
	\subfigcapskip=-5pt %设置子图与子标题之间的距离
	\subfigure[TX Array]{
		\includegraphics[width=0.3\linewidth]{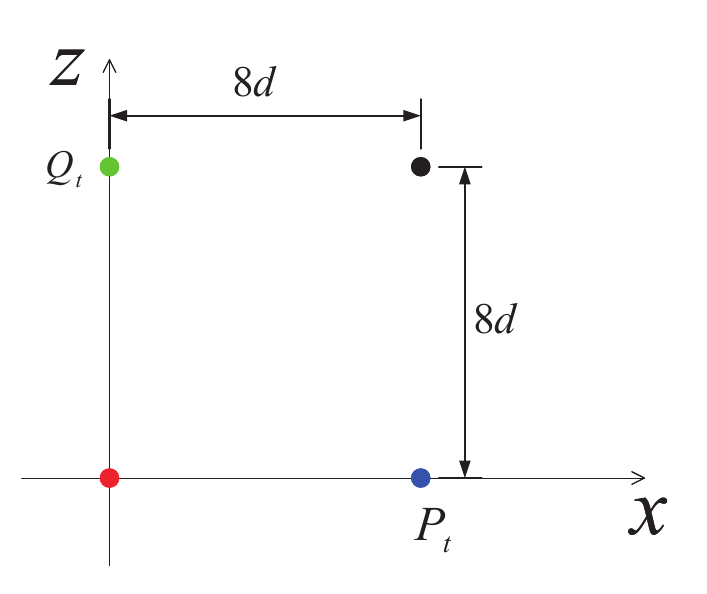}}
	\subfigure[RX Array]{
		\includegraphics[width=0.3\linewidth]{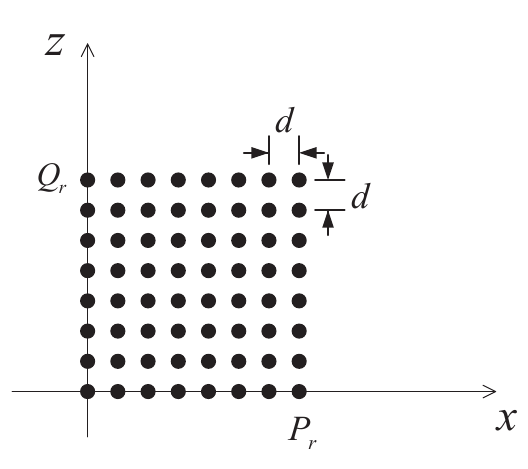}}
	\subfigure[Virtual Rx Array]{
		\includegraphics[width=0.3\linewidth]{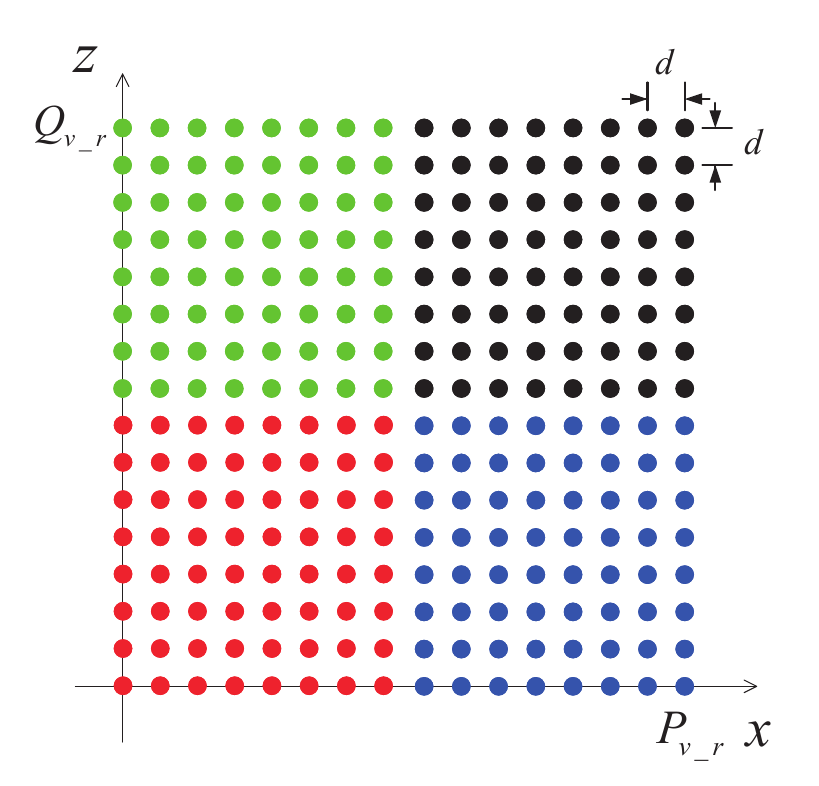}}
	\caption{MIMO array design with the transmitting array in (a), the receiving array in (b) and the corresponding Virtual Rx (VRx) array in (c).}
	\label{fig:Array}
\end{figure}

In order to simultaneously measure the pitch and azimuth angles of DoA and obtain higher angular resolution, we design transmit-receive UPAs as shown in Fig.~\ref{fig:Array}. The receiving antenna array element spacing  is half wavelength, i.e. , $d={{\lambda }/{2}}$, where ${\lambda }$ represents the signal wavelength, the transmit UPA is a ${2\times 2}$ array with a spacing of ${8d}$, the receive UPA is an ${8\times 8}$ array with a spacing of ${d}$, and the corresponding receive UPA is a ${16\times 16}$ array with a spacing of ${d}$.

\begin{figure}[!htb]
	\centering  %图片全局居中
	\includegraphics[width=0.4\linewidth]{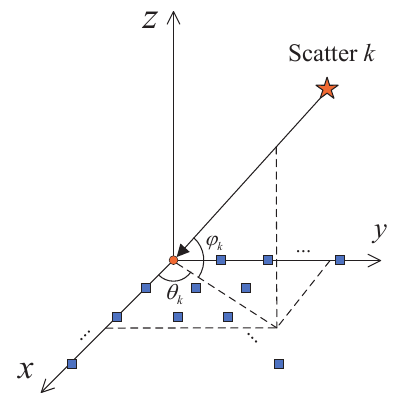}
	\caption{Schematic diagrams of the geometric relationship between the target and the UPAs.}
	\label{fig:position}
\end{figure}

The geometric relationship between the $k$th scattering point and BS is shown in Fig.~\ref{fig:position}, with the azimuth and pitch angles noted as (${{\theta }_{k}}$, ${{\varphi }_{k}}$). The antenna array element in the virtual receiving array is denoted as ${{A}_{v\_r}}({{p}_{v\_r}},{{q}_{v\_r}})$, and the reference antenna array element is ${{A}_{v\_r}}(1,1)$, then the phase difference between ${{A}_{v\_r}}({{p}_{v\_r}},{{q}_{v\_r}})$ and ${{A}_{v\_r}}(1,1)$ caused by the $k$th scattering point can be expressed by (\ref{eq:position}).\newline 
%\begin{equation}
%    \begin{aligned}
%      & \Delta {{\phi }_{v\_r}}{{\left( {{p}_{v\_r}},{{q}_{v\_r}} \right)}_{\left( {{\theta }_{k}},{{\varphi }_{k}} \right)}} \\ 
%     & =\left\{ \begin{aligned}
%      & {{e}^{-j2\pi \frac{d\left[ \left( {{p}_{v\_r}}-1 \right)\cos {{\theta }_{k}}+\left( {{q}_{v\_r}}-1 \right)\sin {{\theta }_{k}} \right]\cos {{\varphi }_{k}}}{\lambda }}}, \\ 
%     & {{\theta }_{k}}<{{90}^{{}^\circ }},{{\varphi }_{k}}<{{90}^{{}^\circ }} \\ 
%     & {{e}^{j2\pi \frac{d\left[ \left( {{p}_{v\_r}}-1 \right)\cos \left( \pi -{{\theta }_{k}} \right)-\left( {{q}_{v\_r}}-1 \right)\sin \left( \pi -{{\theta }_{k}} \right) \right]\cos {{\varphi }_{k}}}{\lambda }}}, \\ 
%     & {{\theta }_{k}}>{{90}^{{}^\circ }},{{\varphi }_{k}}<{{90}^{{}^\circ }} \\ 
%     & {{e}^{-j2\pi \frac{d\left[ \left( {{p}_{v\_r}}-1 \right)\cos {{\theta }_{k}}-\left( {{q}_{v\_r}}-1 \right)\sin {{\theta }_{k}} \right]\cos \left( \pi -{{\varphi }_{k}} \right)}{\lambda }}}, \\ 
%     & {{\theta }_{k}}<{{90}^{{}^\circ }},{{\varphi }_{k}}>{{90}^{{}^\circ }} \\ 
%     & {{e}^{j2\pi \frac{d\left[ \left( {{p}_{v\_r}}-1 \right)\cos \left( \pi -{{\theta }_{k}} \right)+\left( {{q}_{v\_r}}-1 \right)\sin \left( \pi -{{\theta }_{k}} \right) \right]\cos \left( \pi -{{\varphi }_{k}} \right)}{\lambda }}}, \\ 
%     & {{\theta }_{k}}>{{90}^{{}^\circ }},{{\varphi }_{k}}>{{90}^{{}^\circ }} \\ 
%    \end{aligned} \right. \\ 
%    \end{aligned}
%    \label{eq:position}
%\end{equation}

\begin{equation}
	\footnotesize
	\begin{aligned}
		& \Delta {{\phi }_{v\_r}}{{\left( {{p}_{v\_r}},{{q}_{v\_r}} \right)}_{\left( {{\theta }_{k}},{{\varphi }_{k}} \right)}} \\
		&={{e}^{-\xi j2\pi \frac{d\left[ \left( {{p}_{v\_r}}-1 \right)\cos {{\theta }_{k}}+\psi \left( {{q}_{v\_r}}-1 \right)\sin {{\theta }_{k}} \right]\cos {{\varphi }_{k}}}{\lambda }}} \\ 
		& \left\{ \begin{aligned}
			& \xi =1\ \psi =1,{{\theta }_{k}}<{{90}^{{}^\circ }},{{\varphi }_{k}}<{{90}^{{}^\circ }} \\ 
			& \xi =-1\ \psi =-1,{{\theta }_{k}}>{{90}^{{}^\circ }},{{\varphi }_{k}}<{{90}^{{}^\circ }} \\ 
			& \xi =1\ \psi =-1,{{\theta }_{k}}<{{90}^{{}^\circ }},{{\varphi }_{k}}>{{90}^{{}^\circ }} \\ 
			& \xi =-1\ \psi =1,{{\theta }_{k}}>{{90}^{{}^\circ }},{{\varphi }_{k}}>{{90}^{{}^\circ }} \\ 
		\end{aligned} \right. \\ 
	\end{aligned}	
	\label{eq:position}
\end{equation}

As shown in (\ref{eq:Ak_matrix}), the VRx array phase difference matrix ${{\mathbf{A}}_{k}}$ corresponding to the $k$th target is obtained from (\ref{eq:position}).

\begin{equation}
	\centering  %全局居中
	%\footnotesize
		{{\mathbf{A}}_{k}}\left( {{p}_{v\_r}},{{q}_{v\_r}} \right)=\!\Delta {{\phi }_{v\_r}}{{\left( {{p}_{v\_r}},{{q}_{v\_r}} \right)}_{\left( {{\theta }_{k}},{{\varphi }_{k}} \right)}} 
%	{{\mathbf{A}}_{k}}=\left( \begin{matrix}
%		1 \!&\!\cdots \!&\!\Delta {{\phi }_{v\_r}}{{\left( {{p}_{v\_r}},1 \right)}_{\left( {{\theta }_{k}},{{\varphi }_{k}} \right)}}  \\
%		\vdots\!&\!\ddots\!&\!\vdots   \\
%		\Delta {{\phi }_{v\_r}}{{\left( 1,{{q}_{v\_r}} \right)}_{\left( {{\theta }_{k}},{{\varphi }_{k}} \right)}} \!&\!\cdots\!&\!\Delta {{\phi }_{v\_r}}{{\left( {{p}_{v\_r}},{{q}_{v\_r}} \right)}_{\left( {{\theta }_{k}},{{\varphi }_{k}} \right)}}  \\
%	\end{matrix} \right)
	\label{eq:Ak_matrix}
\end{equation}

%\begin{figure}[!htb]
%	\centering  %图片全局居中
%	\subfigbottomskip=2pt %两行子图之间的行间距
%	\subfigcapskip=-5pt %设置子图与子标题之间的距离
%	\subfigure[${{\theta }_{k}}<90{}^\circ $,${{\varphi }_{k}}<90{}^\circ $.]{
%		\includegraphics[width=0.35\linewidth]{figure/目标方位示意图_1.eps}}
%	\subfigure[${{\theta }_{k}}>90{}^\circ $,${{\varphi }_{k}}<90{}^\circ $.]{
%		\includegraphics[width=0.35\linewidth]{figure/目标方位示意图_2.eps}}
%	  \\
%	\subfigure[${{\theta }_{k}}<90{}^\circ $,${{\varphi }_{k}}>90{}^\circ $.]{
%		\includegraphics[width=0.35\linewidth]{figure/目标方位示意图_3.eps}}
%  	\subfigure[${{\theta }_{k}}>90{}^\circ $,${{\varphi }_{k}}>90{}^\circ $.]{
%		\includegraphics[width=0.35\linewidth]{figure/目标方位示意图_4.eps}}
%%        \\
%%    \subfigure[Suspension method of UPAs.]{
%%    \includegraphics[width=2.5cm,keepaspectratio]{figure/目标方位示意图_5.eps}}
%	\caption{(a)-(d) are schematic diagrams of the geometric relationship between the target and the UPAs.}
%%		, and (e) is the antenna suspension method of the constructed scenes in this paper.}
%    \label{fig:position}
%\end{figure}

\subsection{DL ISAC Signal Model}
\begin{figure}[!htb]
	\centering  %图片全局居中
	\subfigbottomskip=2pt %两行子图之间的行间距
	\subfigcapskip=-5pt %设置子图与子标题之间的距离
	\subfigure[Comb4 with 4 symbols.]{
		\includegraphics[width=3.6cm,keepaspectratio]{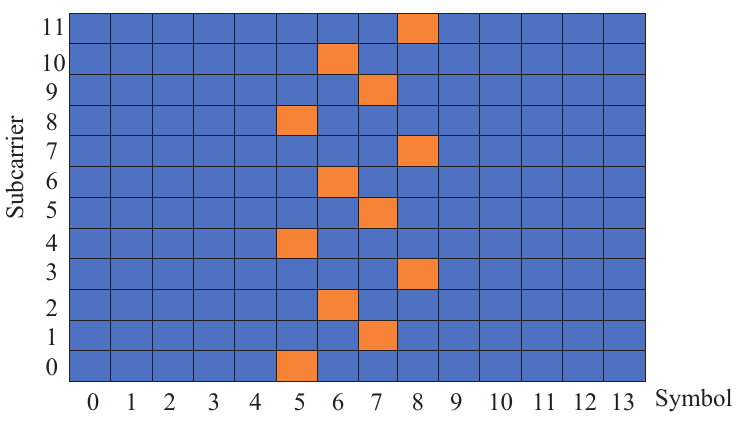}}
	\subfigure[PRS mapped OFDM signal model.]{
		\includegraphics[width=4.8cm,keepaspectratio]{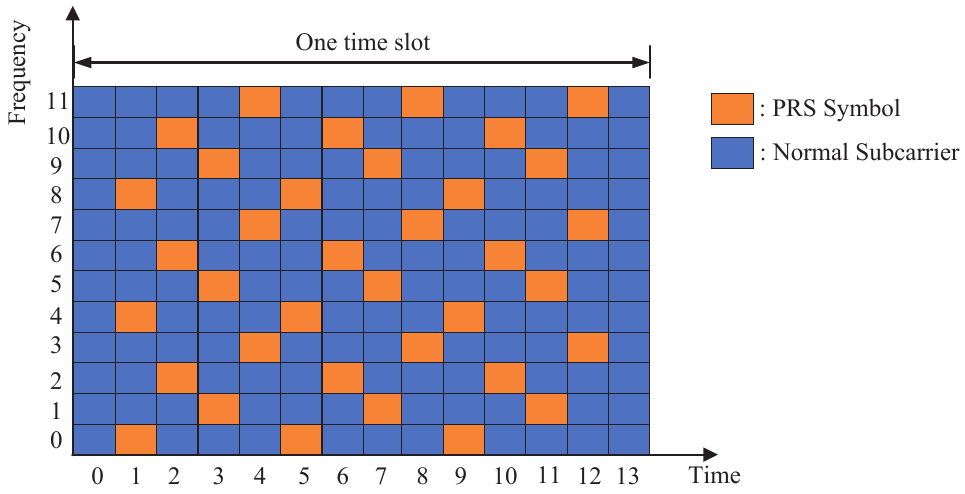}}
	\caption{PRS time-frequency structure schematic.}
	\label{fig:PRS}
\end{figure}

As shown in Fig.~\ref{fig:PRS} (a) we use the structure of Comb4 with 4 symbols and the PRS mapped OFDM signal model is shown in Fig.~\ref{fig:PRS} (b)~\cite{wei20225g}.In a time slot, PRS occupies ${{N}_{PRS_t}}$ OFDM symbols for the time domain where ${\text{PRS}_t}$ is the set of symbols carrying PRS, and PRS occupies${{N}_{\text{PRS}_f}}$ subcarriers at certain spacings for the frequency domain where ${\text{PRS}_f}$ is the set of subcarries carrying PRS.

The continuous time domain OFDM signal can be described as
\begin{equation}
\footnotesize
    y\left ( t \right )=\rm \sum_{m=0}^{N_{sym}-1 } \rm \sum_{n=0}^{N_c-1}{{s}_{Tx}}\left ( n,m \right )e^{j2\pi f_{n}t}{\rm rect}\left ( \frac{t-mT_{{\rm OFDM}} }{T_{{\rm OFDM}}}  \right )
\label{eq:OFDM}
\end{equation}
where ${s_{Tx}\left ( n,m \right )}$ represents the modulated OFDM symbol in the $n$th subcarrier of $m$th OFDM symbol, $N_{sym}$ is the number of OFDM symbols and $N_{c}$ is the number of subcarriers. $T_{\text{OFDM}}$ is the total duration of the OFDM symbol which satisfies ${{T}_{\text{OFDM}}}=T+{{T}_{\text{CP}}}$, where $T$ is the effective OFDM symbol duration and ${T}_{\text{CP}}$ is the cyclic prefix duration. $\Delta f={1}/{T}\;$ is the frequency interval of subcarriers, $f_n$ is the frequency of the $n$th subcarriers that carrying the modulation symbol, $\text{rect}\left( \frac{t}{{{T}_{\text{OFDM}}}} \right)$ is the rectangular function which is equal to 1 for $0\le t\le {{T}_{\text{OFDM}}}$ and 0 for otherwise.

%\addtolength{\topmargin}{0.04in}

When $n\in {\text{PRS}_{f}},m\in {\text{PRS}_{t}}$, ${s\left ( n,m \right )}$ takes the GOLD sequence value, otherwise ${s\left ( n,m \right )}$ carries random communication information.The PRS occupies the 2nd-13th OFDM symbols in a time slot, and we insert a set of PRS every 4th-time slot. The frequency of the subcarriers carrying PRS symbols $f_n$ satisfies (\ref{eq:f_n}) when ${m\in {\text{PRS}_{t}}}$, where $K_{\text{Comb}}^{\text{PRS}}$ is the comb size of PRS, $n_0$ is the index of
the first subcarrier carrying PRS.
\begin{equation}
    {{f}_{n}}=\left( n\times K_{\text{Comb}}^{\text{PRS}}+{{n}_{0}} \right)\Delta f,n=0,\ldots ,{{N}_{\text{PR}{{\text{S}}_{f}}}}-1
    \label{eq:f_n}
\end{equation}
The designed DL ISAC imaging signal satisfies $K_{\text{Comb}}^{\text{PRS}}$ and ${{n}_{0}}\in [0,2,1,3]$ in turn. The relation between ${{n}_{0}}$
and $m$ is shown in (\ref{eq:n0}), where mod refers to a modulo operation.
\begin{equation}
    {{n}_{0}}=\frac{m\bmod K_{\text{Comb}}^{\text{PRS}}}{2}+\frac{3}{4}\left[ 1-{{\left( -1 \right)}^{m\bmod K_{\text{Comb}}^{\text{PRS}}}} \right]
    \label{eq:n0}
\end{equation}

\subsection{ISAC Received Signal Model}
The received signal can be described as (\ref{eq:OFDM_reflected}) when the OFDM signal shown in (\ref{eq:OFDM}) is reflected by the $k$th scatter. $R_k$ and ${{f}_{d}}\left( k \right)$ is the range and doppler shift of kth scatter, respectively. $G_k$ represents the attenuation factor associated with the path loss, radar cross section (RCS) of the kth scatter.
\begin{equation}
	\begin{aligned}
		& y_{n,m,k}^{Rx}\left( t \right)={{G}_{k}}\sum\limits_{m=0}^{{{N}_{sym}}-1}{{{e}^{j2\pi {{{f}_{d}}\left( k \right)}t}}\sum\limits_{n=0}^{{{N}_{c}}-1}{{{s}_{Rx}}\left( n,m,k \right)}}\times  \\ 
		& {{e}^{j2\pi {{f}_{n}}\left( t-\frac{2{{R}_{k}}}{c} \right)}}\times \text{rect}\left( \frac{t-m{{T}_{\text{OFDM}}}-\frac{2{{R}_{k}}}{c}}{{{T}_{\text{OFDM}}}} \right)  
	\end{aligned}
    \label{eq:OFDM_reflected}
\end{equation}
The relationship between the received modulation symbols ${{{s}_{Rx}}\left( n,m,k \right)}$ and the transmitted modulation symbol ${{{s}_{Tx}}\left( n,m\right)}$ can be described by (\ref{eq:Rx_Tx_reletionship}) from (\ref{eq:OFDM}).
\begin{equation}
	{{s}_{Rx}}\left( n,m,k \right)={{G}_{k}}{{s}_{Tx}}\left( n,m \right){{e}^{-j2\pi {{f}_{n}}\frac{2{{R}_{k}}}{c}}}{{e}^{j2\pi {{{f}_{d}}\left( k \right)mT_{\text{OFDM}}}}}
    \label{eq:Rx_Tx_reletionship}
\end{equation}

Dividing the received modulation symbols ${{{s}_{Rx}}\left( n,m,k \right)}$ and the transmitted modulation symbol ${{{s}_{Tx}}\left( n,m\right)}$,  the following matrix  ${{{s}_{g}}\left( n,m,k \right)}$ is obtained.
\begin{equation}
	{{s}_{g}}\left( n,m,k \right)=\frac{{{s}_{Rx}}\left( n,m,k \right)}{{{s}_{Tx}}\left( n,m \right)}={{G}_{k}}\left( {{{\vec{k}}}_{r}}\otimes {{{\vec{k}}}_{d}} \right)
	\label{eq:matrix_info}
\end{equation}
where ${{\vec{k}}_{r}}=\left( 0,{{e}^{-j2\pi \Delta f\frac{2{{R}_{k}}}{c}}},\cdots ,{{e}^{-j2\pi \left( {{N}_{c}}-1 \right)\Delta f\frac{2{{R}_{k}}}{c}}} \right)$ and ${{\vec{k}}_{d}}=\left( 0,{{e}^{j2\pi {{T}_{\text{OFDM}}}{{f}_{d}}\left( k \right)}},\cdots ,{{e}^{j2\pi ({{N}_{sym}}-1){{T}_{\text{OFDM}}}{{f}_{d}}\left( k \right)}} \right)$ are the two vectors carrying the range and the doppler information. $\otimes$ refers to a dyadic product~\cite{sturm2011waveform}.

The detection information matrix ${\mathbf{A}}_{S}$ of all array elements in the corresponding virtual receiving antenna array can be obtained by combining (\ref{eq:Ak_matrix}) and (\ref{eq:matrix_info}) as shown in (\ref{eq:Array_signal_sumk}).
\begin{equation}
%	\footnotesize
%	\begin{aligned}
	%	& 
		{\mathbf{A}}_{S}=\sum\limits_{k}{{{s}_{g}}\left( n,m,k \right){{\mathbf{A}}_{k}}} %\\ 
%		& =\left( \begin{matrix}
%			\rm \sum\limits_{k}{{{s}_{g}}} \rm & \rm\cdots \rm & \rm \sum\limits_{k}{{{s}_{g}}\Delta {{\phi }_{v\_r}}{{\left( {{p}_{v\_r}},1 \right)}_{\left( {{\theta }_{k}},{{\varphi }_{k}} \right)}}}  \\
%			\vdots  \rm & \rm \ddots \rm  & \rm \vdots   \\
%			\rm \sum\limits_{k}{{{s}_{g}}\Delta {{\phi }_{v\_r}}{{\left( 1,{{q}_{v\_r}} \right)}_{\left( {{\theta }_{k}},{{\varphi }_{k}} \right)}}} \rm & \rm \cdots \rm  & \rm \sum\limits_{k}{{{s}_{g}}\Delta {{\phi }_{v\_r}}{{\left( {{p}_{v\_r}},{{q}_{v\_r}} \right)}_{\left( {{\theta }_{k}},{{\varphi }_{k}} \right)}}}  \\
%		\end{matrix} \right)  
%	\end{aligned}
	\label{eq:Array_signal_sumk}
\end{equation}
The $A_S$ in our simulation work exists in the form of 4D arrays which contain the dimensions of fast time dimension, slow time dimension, horizontal antenna array dimension and vertical antenna array dimension. In actual ISAC 4D imaging system, $A_S$ represents the environmental scatterer reflection signal received by UPA.

\section{Methodology of Signal Processing}
\begin{figure}[!htb]
	\centering
	\includegraphics[width=\linewidth]{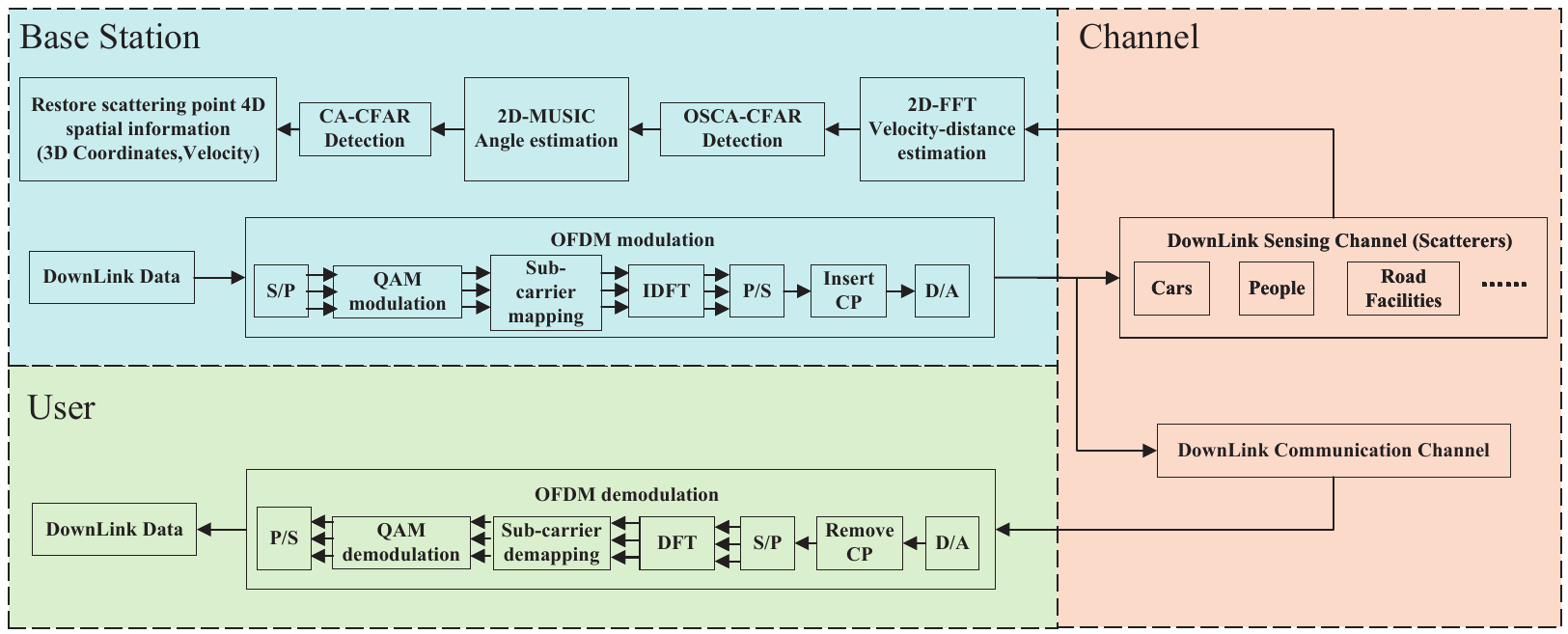}
	\caption{Diagram of the DL ISAC signal processing.}
	\label{fig:system_flow}
\end{figure}
In this section, we demonstrate the sensing and communication processing method as shown in Fig.~\ref{fig:system_flow}. We present the simulation method of ISAC imaging environment, the ISAC imaging sensing processing method, and the ISAC Imaging communication processing method.

\subsection{ISAC Imaging Environment Simulation}
\begin{figure}[!htb]
	\centering
	\includegraphics[width=0.5\linewidth]{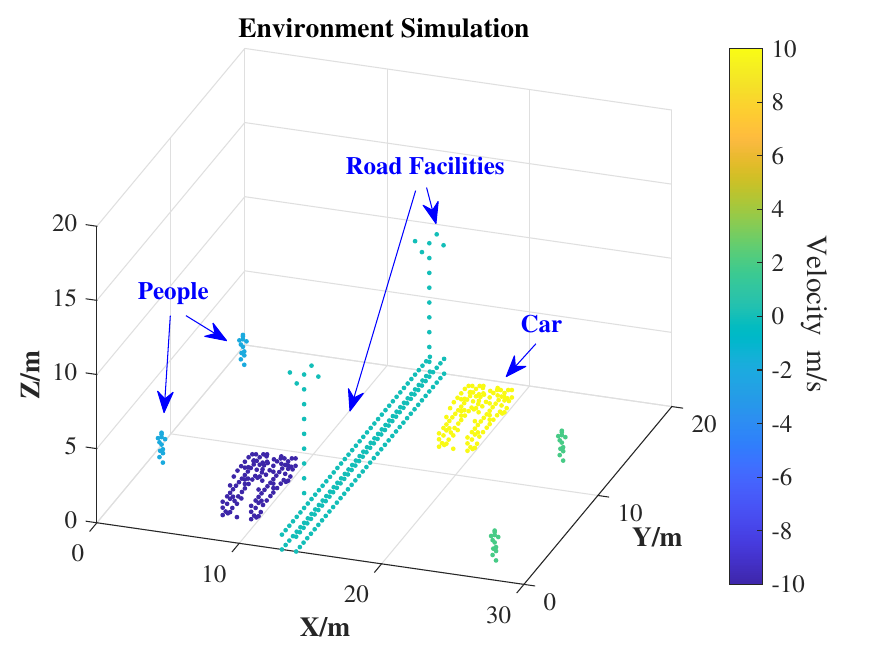}
	\caption{Simulation of environmental scattering points.}
	\label{fig:original_environment}
\end{figure}
As shown in Fig.~\ref{fig:original_environment}, we used scatter points to depict the contours of people, vehicles and road infrastructure in the simulated environment. We also added $\pm 2\text{m}/\text{s}$, $\pm 10\text{m}/\text{s}$ and stationary velocity information for the above three types of targets respectively. In addition, we assumed that the ISAC BS is suspended at ${{P}_{BS}}\left( x,y,z \right)=\left( 14,100,20 \right)$.

\subsection{ISAC Sensing Signal Processing}
In this section we first introduce the range and Doppler detection based on 2D-FFT and OSCA-CFAR algorithms, then introduce the DoAs estimation based on 2D-MUSIC and CA-CFAR algorithms.
\subsubsection{ISAC Imaging Range and Doppler Detection}
The ${{{\vec{k}}}_{r}}$ in (\ref{eq:matrix_info}) is the linear phase shift vector along the fast time axis caused by range $R$. Therefore, the scatters’ range can be estimated by using (\ref{eq:matrix_info}) to initially process the received signal of each antenna array element and then performing the Inverse Discrete Fourier  Transform (IDFT) along the subcarrier dimension.
\begin{equation}
	\begin{aligned}
		& R\left( \alpha  \right)=\text{IDFT}\left[ {{k}_{r}}\left( n \right) \right]=\frac{1}{{{N}_{c}}}\sum\limits_{n=0}^{{{N}_{c}}-1}{{{k}_{r}}\left( n \right){{e}^{jn\alpha \frac{2\pi }{{{N}_{c}}}}}} \\ 
		& =\frac{1}{{{N}_{c}}}\sum\limits_{n=0}^{{{N}_{c}}-1}{{{e}^{-j2\pi n\Delta f\frac{2R}{c}}}{{e}^{jn\alpha \frac{2\pi }{{{N}_{c}}}}}},\alpha =0,\ldots ,{{N}_{c}}-1  
	\end{aligned}
	\label{eq:kr_fft}
\end{equation}
where $R\left( \alpha  \right)$ obtains peak value when $\alpha =\left\lfloor \frac{2R\Delta f{{N}_{c}}}{c} \right\rfloor $.The ${{{\vec{k}}}_{d}}$ in (\ref{eq:matrix_info}) is the linear phase shift vector along the slow time axis caused by velocity $v_k$. Similarly $v_k$ can be estimated by performing a Discrete Fourier Transform (DFT) along the OFDM symbol dimension.
\begin{equation}
	\begin{aligned}
		\rm &\rm  v\left( \beta  \right)=\text{DFT}\rm \left[ {{k}_{d}}\left( m \right) \right]=\sum\limits_{m=0}^{{{N}_{sym}}-1}{{{k}_{d}}\left( m \right){{e}^{-jm\beta \frac{2\pi }{{{N}_{sym}}}}}} \\ 
		\rm &\rm  =\rm \sum\limits_{m=0}^{{{N}_{sym}}-1}\rm {{{e}^{j2\pi m{{T}_{\text{OFDM}}}\frac{2{{v}_{k}}{{f}_{c}}}{c}}}{{e}^{-jm\beta \frac{2\pi }{{{N}_{sym}}}}}},\beta =0,\ldots ,{{N}_{sym}}-1 \\ 
	\end{aligned}
	\label{eq:kd_fft}
\end{equation}
where ${v\left( \beta  \right)}$obtains peak value when $\beta$ in (\ref{eq:kd_fft}) satisfies $\beta =\left\lfloor \frac{2{{v}_{k}}{{f}_{c}}{{T}_{\text{OFDM}}}{{N}_{sym}}}{c} \right\rfloor $. 

\begin{figure}[!htb]
	\centering  %图片全局居中
	\subfigbottomskip=2pt %两行子图之间的行间距
	\subfigcapskip=-5pt %设置子图与子标题之间的距离
	\subfigure[2D OSCA-CFAR.]{
		\includegraphics[width=0.35\linewidth]{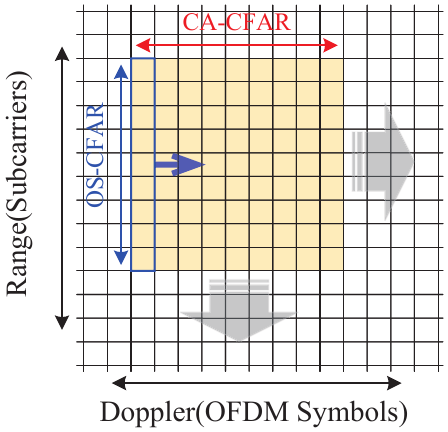}}
	\subfigure[2D CA-CFAR.]{
		\includegraphics[width=0.35\linewidth]{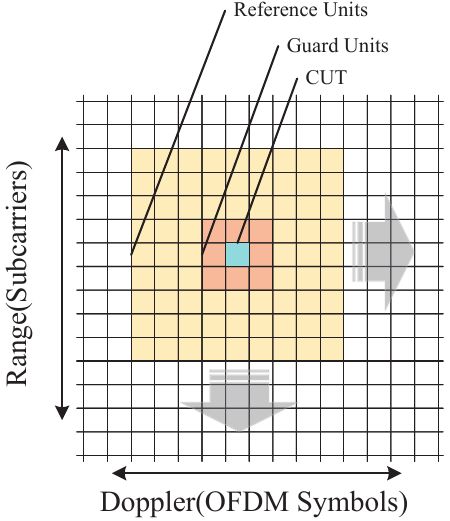}}
	\\
	\subfigure[Threshold.]{
		\includegraphics[width=3.5cm,keepaspectratio]{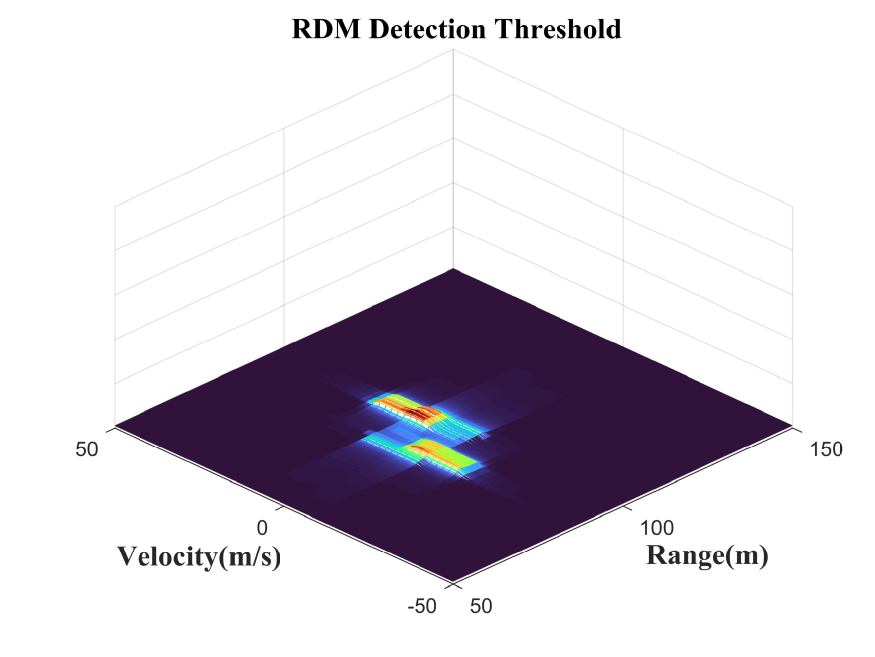}}
	\subfigure[RDM.]{
		\includegraphics[width=3.5cm,keepaspectratio]{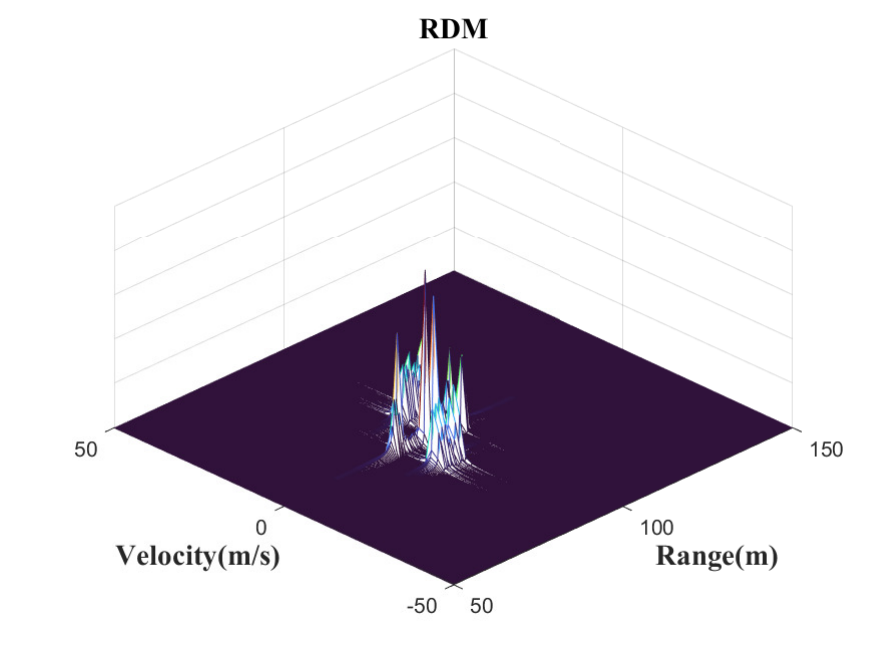}}
	\\
	\subfigure[Threshold.]{
		\includegraphics[width=3.5cm,keepaspectratio]{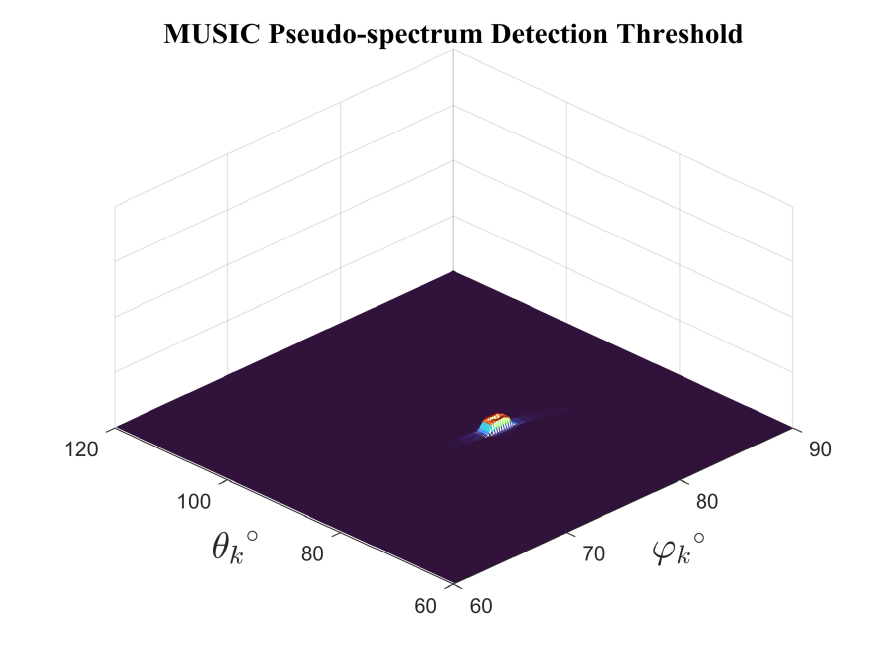}}
	\subfigure[MUSIC Pseudo-spectrum.]{
		\includegraphics[width=3.5cm,keepaspectratio]{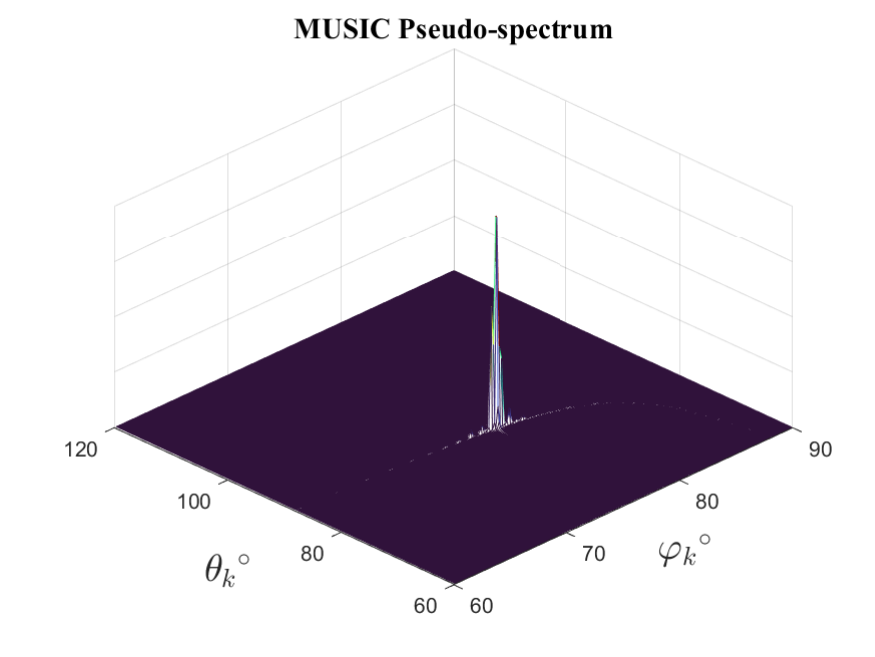}}
	\caption{(a)-(b) are the sliding reference window design for OSCA-CFAR and CA-CFAR, (c)-(d) are the RDM and its detection threshold and (e)-(f) are the MUSIC Pseudo-spectrum and its detection threshold.}
	\label{fig:CFAR}
\end{figure}

We denote the 2D-FFT processing result of $s_g$ in (\ref{eq:matrix_info}) as $s_{_{g}}^{R,v}$ which is known as RDM, and then we estimate the range and velocity of scattering points using 2D OSCA-CFAR detection in $s_{_{g}}^{R,v}$.As shown in Fig.~\ref{fig:CFAR} (a) we select a reference window of size $9\times9$ within the RDM.

\begin{itemize}
\item Firstly, the 1D OS-CFAR is applied to the sliding reference window by column and the $\gamma$th value of each colum is selected as the noise estimate of this sliding window. We take $\gamma=\left\lfloor 0.75N \right\rfloor $, where $N$ is the number of reference cells.
\item  Then the noise estimation values of the above columns are averaged by row as shown in (\ref{eq:OSCA_CFAR}) as the noise threshold of the Center Detection Unit (CUT) in reference window.
\begin{equation}
	{{\bar{\mu }}_{\left( \gamma \right)}}=\frac{1}{N}\sum\limits_{n=1}^{N}{{{X}_{\left( \gamma \right),n}}}
	\label{eq:OSCA_CFAR}
\end{equation}
where ${X}_{\left( \gamma \right),n}$ represents the noise reference value selected by OS-CFAR, $N=9$.
\item As shown in (\ref{eq:threshold_factor}), the detection threshold factor ${{T}_{f}}$ of this CUT is only related to the number of reference units $N$ with the specified false alarm probability ${{P}_{fa}}$. Finally, the detection threshold $T={{T}_{f}}\times {{\bar{\mu }}_{\left( l \right)}}$ for a single CUT can be obtained, and the 2D adaptive detection threshold can be obtained by applying the above operations to all CUTs in the RDM as shown in Fig. ~\ref{fig:CFAR} (d)-(e).
\begin{equation}
	{{T}_{f}}={{\left( {{P}_{fa}} \right)}^{-\frac{1}{N}}}-1
	\label{eq:threshold_factor}
\end{equation}

%\begin{figure}[!htb]
%	%\setlength {\abovecaptionskip} {0.cm} 
%	%\setlength {\belowcaptionskip} {-0.cm}
%	\centering  %图片全局居中
%	\subfigbottomskip=2pt %两行子图之间的行间距
%	\subfigcapskip=-5pt %设置子图与子标题之间的距离
%	\subfigure[Threshold.]{
%		\includegraphics[width=3.5cm,keepaspectratio]{figure/RDM_threshold.eps}}
%	\subfigure[RDM.]{
%		\includegraphics[width=3.5cm,keepaspectratio]{figure/RDM_result.eps}}
%	\caption{RDM and its detection threshold.}
%	\label{fig:RDM}
%\end{figure}

\end{itemize}

\subsubsection{ISAC Imaging DoAs Estimation}
The peak of the RDM indicates the presence of targets with velocity $v$ and range $R$ but 2D angles are unknown.
%and the linear phase shift of the corresponding $s_{_{g}}^{R,v}$ value due to $R$ and $v$ has been removed~\cite{sturm2011waveform}, which carries only the linear phase shift caused by the wave range difference of the receiving antenna array elements
As shown in (\ref{eq:A_km}), we combine the RDM peaks detected by OSCA-CFAR of all receiving antenna elements into ${{k}_{m}}\left( {{k}_{m}}<k \right)$ manifolds for MUSIC-based DoA estimation .

\begin{equation}
	\footnotesize
%	\begin{aligned}
%		& 
		{{\mathbf{A}}_{{{k}_{m}}}}\left( {{p}_{v\_r}},{{q}_{v\_r}} \right)\rm=\rm\sum\limits_{t}\rm{{{G}_{t}}{{\mathbf{A}}_{t}}} =\sum\limits_{t}\rm {{{G}_{t}}\Delta {{\phi }_{v\_r}}{{\left( {{p}_{v\_r}},{{q}_{v\_r}} \right)}_{\left( {{\theta }_{t}},{{\varphi }_{t}} \right)}}} 
%		\\ 
%		& =\left( \begin{matrix}
%			\sum\limits_{t}{{{G}_{t}}} \rm&\rm  \cdots  \rm&\rm  \sum\limits_{t}{{{G}_{t}}\Delta {{\phi }_{v\_r}}{{\left( {{p}_{v\_r}},1 \right)}_{\left( {{\theta }_{t}},{{\varphi }_{t}} \right)}}}  \\
%			\vdots  \rm & \rm \ddots  \rm &\rm  \vdots   \\
%			\sum\limits_{t}{{{G}_{t}}\Delta {{\phi }_{v\_r}}{{\left( 1,{{q}_{v\_r}} \right)}_{\left( {{\theta }_{t}},{{\varphi }_{t}} \right)}}} \rm & \rm \cdots  \rm &\rm \sum\limits_{t}{{{G}_{t}}\Delta {{\phi }_{v\_r}}{{\left( {{p}_{v\_r}},{{q}_{v\_r}} \right)}_{\left( {{\theta }_{t}},{{\varphi }_{t}} \righ t)}}} \\
%		\end{matrix} \right)  
%	\end{aligned}
	\label{eq:A_km}
\end{equation}

We take ${{\left( {{\mathbf{A}}_{{{k}_{m}}}} \right)}_{1,:}}$ (row where the reference element is located) and ${{\left( {{\mathbf{A}}_{{{k}_{m}}}} \right)}_{:,1}}$ (column where the reference element is located) of ${{\mathbf{A}}_{{{k}_{m}}}}$ to construct the searching manifolds, which are related to the 2D angle ${\mathbf{p}}=\left( \theta ,\varphi  \right)$ of the targets simultaneously, and eventually just multiply the two search results to get the 2D angle information of $t$ targets in ${{\mathbf{A}}_{{{k}_{m}}}}$.Take ${{\left( {{\mathbf{A}}_{{{k}_{m}}}} \right)}_{:,1}}$ as an example to introduce the angle search process.

\begin{figure}[!htb]
	\centering
	\includegraphics[width=0.45\linewidth]{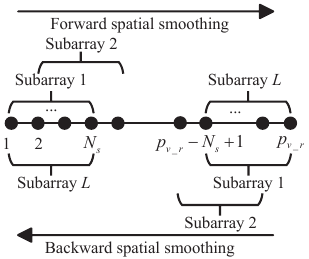}
	\caption{Schematic of the forward and backward space smoothing algorithm.}
	\label{fig:foward_smoothing}
\end{figure}

Since the strong correlation of the received signals, it is necessary to use the MUSIC algorithm based on spatial smoothing. As shown in Fig.~\ref{fig:foward_smoothing}, we can define the forward spatial smoothing matrix ${{\mathbf{R}}_{f}}=\frac{1}{L}\sum\limits_{l=1}^{L}{{\mathbf{R}}_{l}^{f}}$, and similarly the backward spatial smoothing matrix ${{\mathbf{R}}_{b}}=\frac{1}{L}\sum\limits_{l=1}^{L}{{\mathbf{R}}_{l}^{b}}$, where ${{\mathbf{R}}_{l}^{f}}$ and ${{\mathbf{R}}_{l}^{b}}$ are the covariance matrices of the subarray and $L$ is the number of subarray elements. Then use the average of them ${\mathbf{R_X}}=\frac{{{\mathbf{R}}_{f}}+{{\mathbf{R}}_{b}}}{2}$ to replace the original covariance matrix.

%\addtolength{\topmargin}{0.01in}

We can get (\ref{eq:eig}) by applying eigenvalue decomposition to ${\mathbf{R_X}}$.
\begin{equation}
	\left[ {{\mathbf{U}}_{x}},{{\mathbf{\Sigma }}_{x}} \right]=\text{eig}\left( {{\mathbf{{R}_{X}}}} \right)
	\label{eq:eig}
\end{equation}
where ${{\mathbf{\Sigma }}_{x}}$ is the real-valued eigenvalue diagonal matrix and ${{\mathbf{U}}_{x}}$ is the orthogonal eigenmatrix. We obtain the number of targets $k_s$ by performing 1D CA-CFAR detection on ${{\mathbf{\Sigma }}_{x}}$.

Construct ${{U}_{N}}={{U}_{x}}\left( :,1:{{N}_{s}}-{{N}_{x}} \right)$ as the noise subspace basis.Then we can obtain the spatial angular spectrum function as ~\cite{haardt2014subspace}
\begin{equation}
	{{f}_{a}}\left( {\mathbf{p}};{{\mathbf{U}}_{N}} \right)={{\mathbf{a }}^{\text{H}}}\left( {\mathbf{p}} \right){{\mathbf{U}}_{N}}{{\left( {{\mathbf{U}}_{N}} \right)}^{\text{H}}}{\mathbf{a}} \left( {\mathbf{p}} \right)
	\label{eq:spatial_spectrum}
\end{equation}
where  ${\mathbf{p}}=\left( \theta ,\varphi  \right)$ is the 2D angle, ${\mathbf{a}}\left( {\mathbf{p}} \right)$ is given in (\ref{eq:Ak_matrix}).The spatial pseudo-spectrum is represented as ~\cite{haardt2014subspace}
\begin{equation}
	{{S}_{a}^{row}}\left( {\mathbf{p}};{{\mathbf{U}}_{N}} \right)={{\left[ {{\mathbf{a}}^{\text{H}}}\left( {\mathbf{p}} \right){{\mathbf{U}}_{N}}{{\left( {{\mathbf{U}}_{N}} \right)}^{\text{H}}}{\mathbf{a}} \left( {\mathbf{p}} \right) \right]}^{-1}}
	\label{eq:music_spectrum}
\end{equation}

Similarly we can obtain ${{{S}_{a}^{col}}\left( {\mathbf{p}};{{\mathbf{U}}_{N}} \right)}$ by taking the above operation for ${{\left( {{\mathbf{A}}_{{{k}_{m}}}} \right)}_{1,:}}$. We obtained the estimated DoAs by performing 2D CA-CFAR detection on ${{{S}_{a}}\left( {\mathbf{p}};{{\mathbf{U}}_{N}} \right)}={{{S}_{a}^{col}}\left( {\mathbf{p}};{{\mathbf{U}}_{N}} \right)}\odot {{{S}_{a}^{row}}\left( {\mathbf{p}};{{\mathbf{U}}_{N}} \right)}$. $\odot$ represents Hadamard product.

As shown in Fig. ~\ref{fig:CFAR} (b), the noise of the 2D CA-CFAR is determined by averaging the $N_R$ reference cells excluding the protection cell and the CUT.
\begin{equation}
	\bar{\mu }=\frac{1}{{{N}_{R}}}\sum\limits_{n=1}^{{{N}_{R}}}{{{X}_{n}}}
	\label{eq:CA-CFAR}
\end{equation}
where ${{X}_{n}}$ represents the reference unit value.The detection threshold of this CUT can be calculated from $T={{T}_{f}}\times {{\bar{\mu }}}$ where ${{T}_{f}}$ is given by (\ref{eq:threshold_factor}), and the 2D adaptive detection threshold can be obtained by applying the above operations to all CUTs in ${{{S}_{a}}\left( {\mathbf{p}};{{\mathbf{U}}_{N}} \right)}$ as shown in Fig.~\ref{fig:CFAR} (e)-(f). 
%\begin{figure}[!htb]
%	%\setlength {\abovecaptionskip} {0.cm} 
%	%\setlength {\belowcaptionskip} {-0.cm}
%	\centering  %图片全局居中
%	\subfigbottomskip=2pt %两行子图之间的行间距
%	\subfigcapskip=-5pt %设置子图与子标题之间的距离
%	\subfigure[Threshold.]{
%		\includegraphics[width=3.5cm,keepaspectratio]{figure/MUSIC_threshold.eps}}
%	\subfigure[MUSIC Pseudo-spectrum.]{
%		\includegraphics[width=3.5cm,keepaspectratio]{figure/MUSIC_result.eps}}
%	\caption{MUSIC Pseudo-spectrum and its detection threshold.}
%	\label{fig:MUSIC_spectrum}
%\end{figure}

%\subsubsection{4D Point Cloud Reconstruction Method}
%If a scatter is detected as ${{I}_{k}}\left( {{R}_{k}},{{v}_{k}},{{\theta }_{k}},{{\varphi }_{k}} \right)$, the 4D point cloud information ${{P}_{k}}\left( {{x}_{k}},{{y}_{k}},{{z}_{k}},{{v}_{k}} \right)$ can be reconstructed according to (\ref{eq:4D_reconstructed}).
%\begin{equation}
%	\footnotesize
%	\left\{ \begin{aligned}
%		& {{P}_{k}}\left( {{x}_{k}} \right)={{P}_{BS}}\left( x \right)+{{I}_{k}}\left( {{R}_{k}} \right)\sin \left[ {{I}_{k}}\left( {{\varphi }_{k}} \right) \right]\cos \left[ {{I}_{k}}\left( {{\theta }_{k}} \right) \right] \\ 
%		& {{P}_{k}}\left( {{y}_{k}} \right)={{P}_{BS}}\left( y \right)-{{I}_{k}}\left( {{R}_{k}} \right)\sin \left[ {{I}_{k}}\left( {{\varphi }_{k}} \right) \right]\sin \left[ {{I}_{k}}\left( {{\theta }_{k}} \right) \right] \\ 
%		& {{P}_{k}}\left( {{z}_{k}} \right)={{P}_{BS}}\left( y \right)-{{I}_{k}}\left( {{R}_{k}} \right)\cos \left[ {{I}_{k}}\left( {{\varphi }_{k}} \right) \right] \\ 
%	\end{aligned} \right.
%	\label{eq:4D_reconstructed}
%\end{equation}

The overall flow is shown in Algorithm.~\ref{alg:ISP}. Besides, we use the echoes of DL communication signals in the 5G mmWave band to realize the ISAC imaging without occupying additional communication resources. Therefore, the DL communication in our ISAC system is the same as the traditional communication system.

\begin{algorithm}[!htb]
	\caption{ISAC Imaging Signal Processing}
	\label{alg:ISP}
	\renewcommand{\algorithmicrequire}{\textbf{Input:}}
	\renewcommand{\algorithmicensure}{\textbf{Output:}}
	\begin{algorithmic}[1]
		\REQUIRE ISAC Received Signal $y_{n,m}^{Rx}\left( t \right)$ %%input
		\ENSURE Reconstructed 4D scattering point cloud    %%output
		
		\STATE  Get the detection information matrix  ${\mathbf{A}}_{S}$ 
		\STATE	Obtaining RDM using 2D-FFT algorithm 
		\STATE	Obtaining $v$ and $R$ by detecting RDM with OSCA-CFAR \vspace{-0.4cm}
		\FOR{All detected RDM cells $k \in [1,k_m]$}
		\STATE	Constructing DoAs detection manifold ${{\mathbf{A}}_{{{k}_{m}}}}$ 
		\STATE	Decorrelation by using spatial smoothing algorithm 
		\STATE	Get the MUSIC search pseudo-spectrum 
		\STATE	Detecting DoAs with CA-CFAR 
		\ENDFOR
		\STATE	Reduction of detection results to 4D point cloud 
		
		\RETURN ISAC imaging results
	\end{algorithmic}
\end{algorithm}

\section{Numerical and Simulation Results}
In this section, we firstly present our simulation parameters, then compare with traditional 4D-FFT imaging algorithm~\cite{santra2020ambiguity} to demonstrate the superiority of our proposed algorithm. Then the metrics we proposed for evaluating the imaging performance of ISAC is introduced. And  the factors affecting the imaging performance are extensively studied.

\subsection{Simulation Parameter Setting}
%All the simulation parameters we used are shown in Table.~\ref{tab:[parameters]}. 
\begin{table}[!htb]
		\caption{Simulation System Parameters }
		
		\setlength{\tabcolsep}{0.7mm}
		\renewcommand\arraystretch{1.2} %增加表格行距
		\centering
		\begin{tabularx}{0.5\textwidth}{
				>{\centering\arraybackslash}X
				>{\centering\arraybackslash}X
			}
			%保持水平\垂直居中
			\toprule 
			\textbf{Parameter names}&\textbf{Value or Specification} \\ \midrule
			MIMO size & $2\times 2$-Tx/$8\times 8$-Rx/$16\times 16$-VRx \\
			Carrier frequency & 70GHz \\
			Bandwidth & 491.52MHz \\
			Subcarrier spacing & 240kHz  \\
			Subcarrier count & 2048 \\
			OFDM symbol count & 224 \\
			Slot count & 16 \\
			Cyclic Prefix ratio & ${1}/{4}\;$ \\
			Cyclic Suffix ratio & ${1}/{32}\;$ \\
			PRS time slot interval & 4 \\
			PRS structure & Comb4 \\
			Number of OFDM symbols occupied by PRS per slot & 12 \\
			Number of spatially smoothed subarray elements & 8 \\ 
			\bottomrule
		\end{tabularx}
	\label{tab:[parameters]}
\end{table}

\subsection{Qualitative Analysis of Imaging Results}
\begin{figure}[!htb]
	\centering  %图片全局居中
	\subfigbottomskip=2pt %两行子图之间的行间距
	\subfigcapskip=-5pt %设置子图与子标题之间的距离
	\subfigure[FFT, -20dB SNR.]{
		\includegraphics[width=0.31\linewidth]{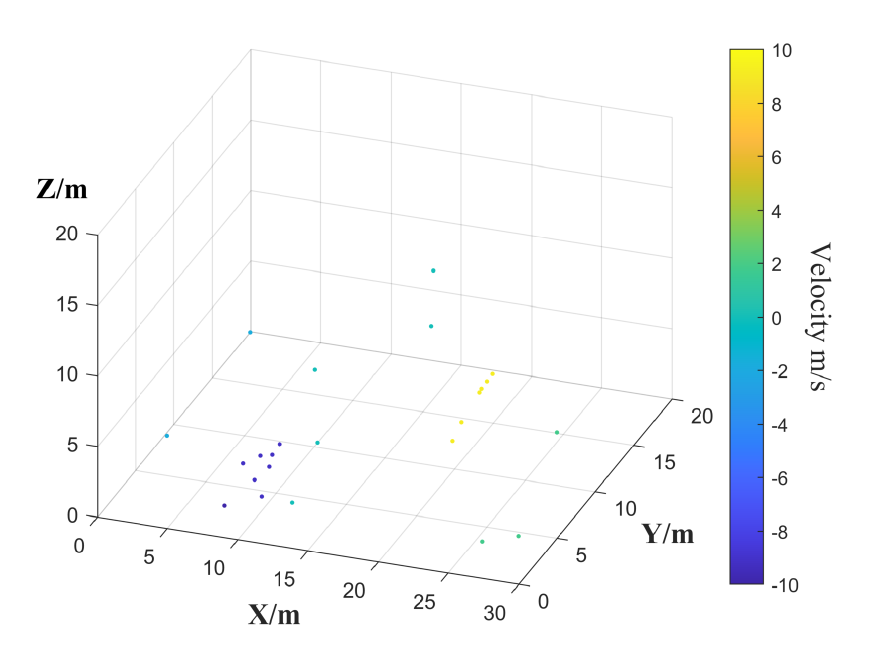}}
	\subfigure[FFT, -5dB SNR]{
		\includegraphics[width=0.31\linewidth]{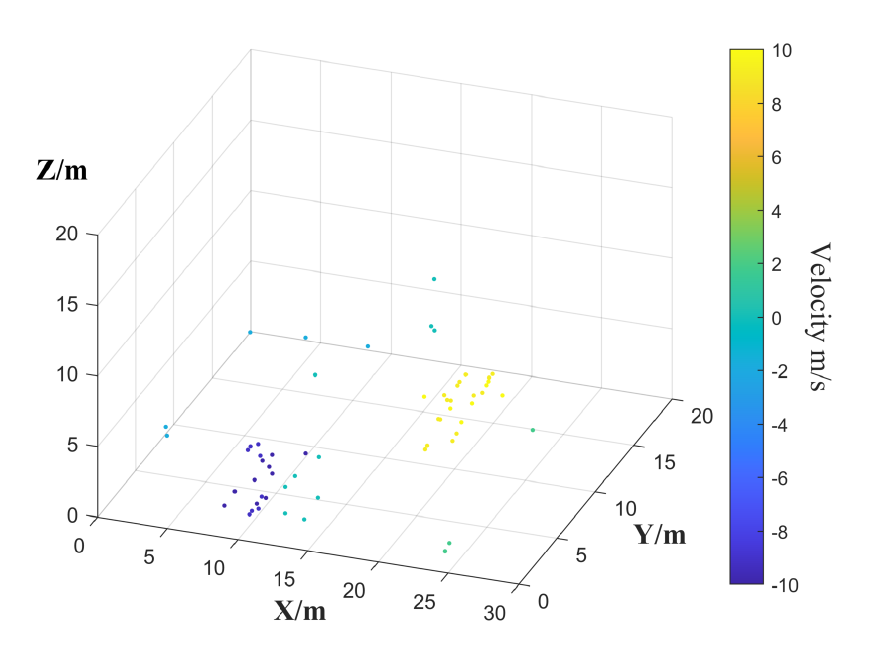}}
	\subfigure[FFT, 10dB SNR]{
		\includegraphics[width=0.31\linewidth]{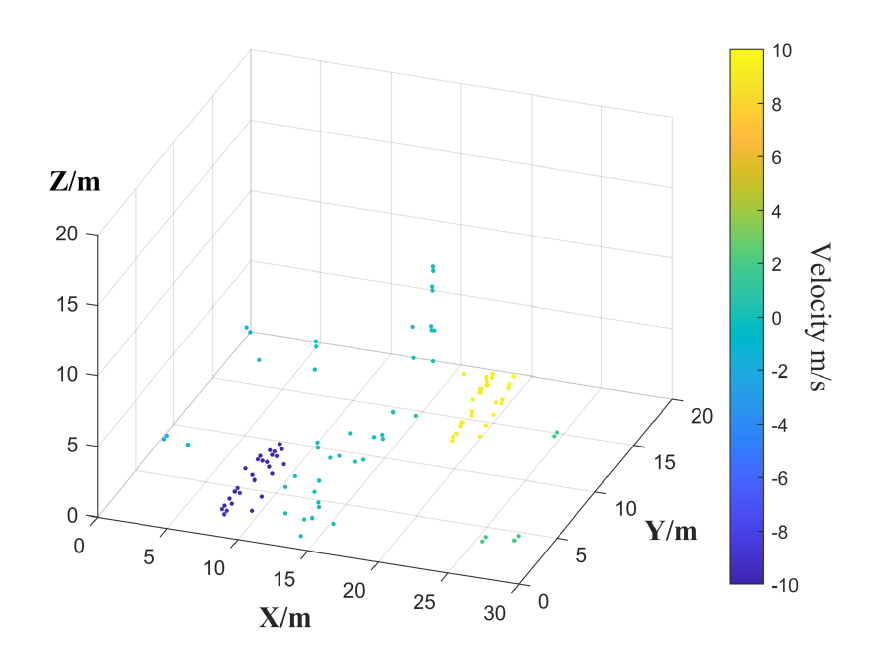}}
	\\
	\subfigure[MUSIC, -20dB SNR]{
		\includegraphics[width=0.31\linewidth]{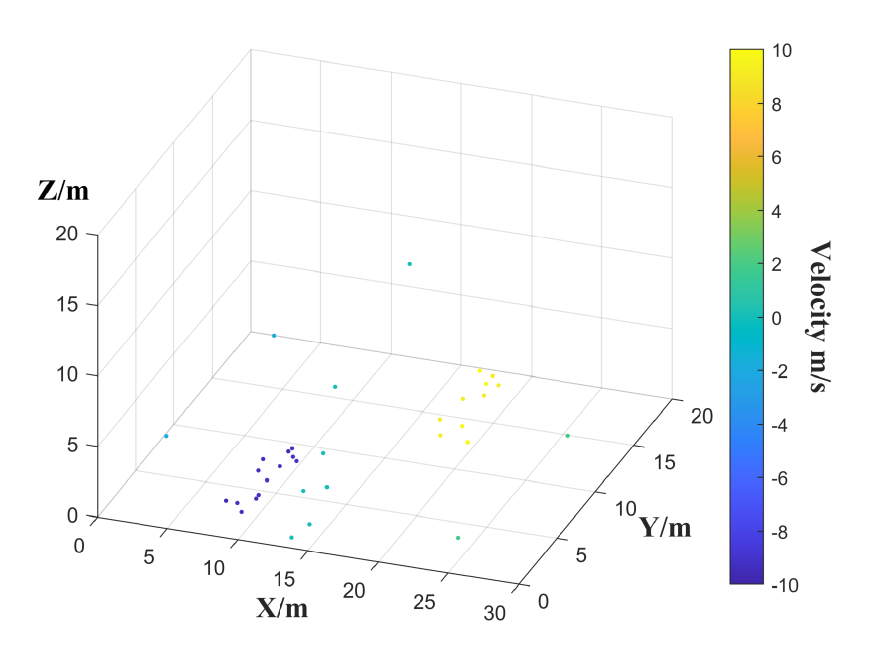}}
	\subfigure[MUSIC, -5dB SNR]{
		\includegraphics[width=0.31\linewidth]{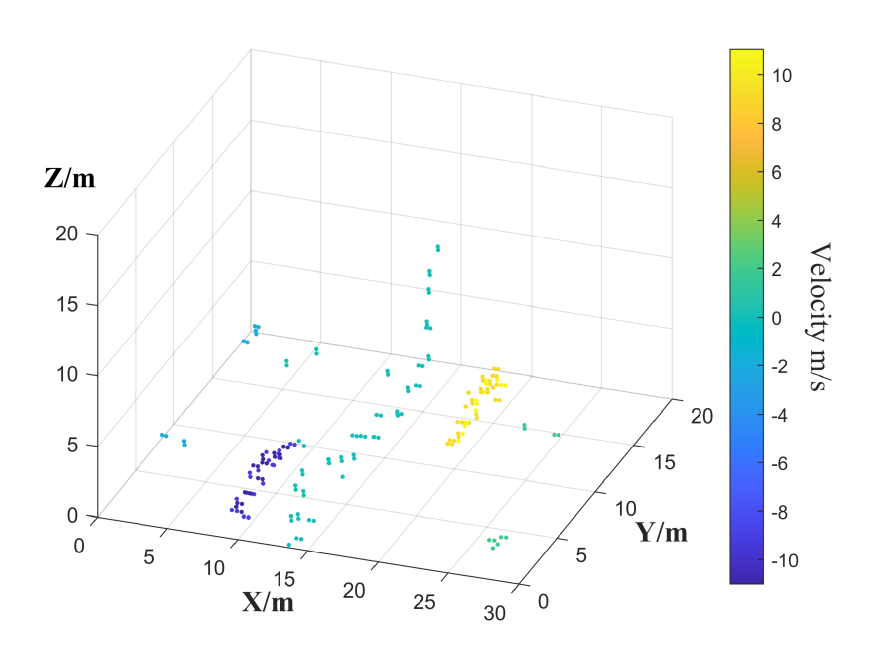}}
	\subfigure[MUSIC, 10dB SNR]{
		\includegraphics[width=0.31\linewidth]{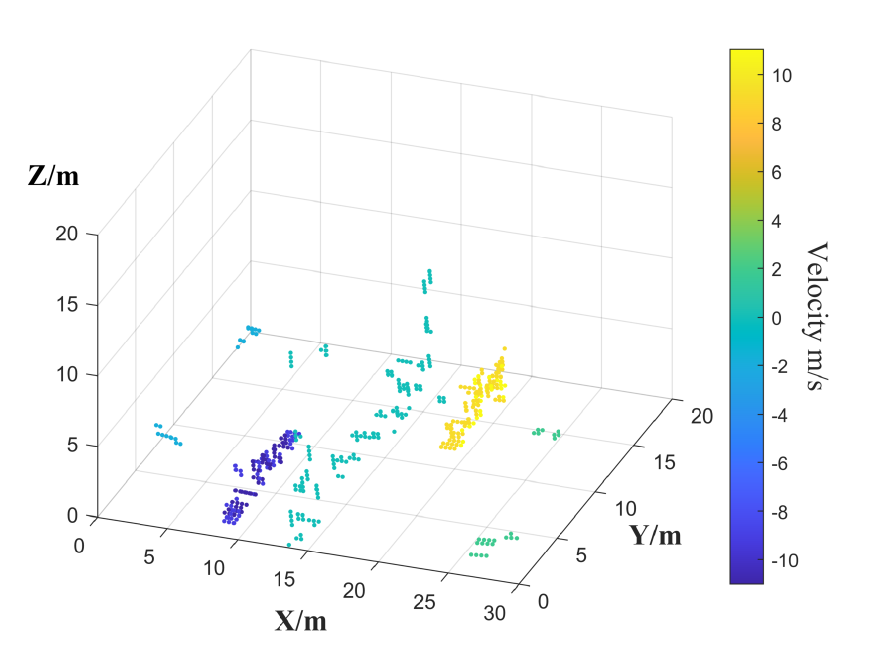}}
	
	\caption{(a)-(c) are the ISAC imaging results under different SNR based on the 4D-FFT algorithm, and (d)-(f) are the ISAC imaging results under different SNR based on the 2D-FFT with 2D-MUSIC algorithm.}
	\label{fig:Imaging_result_compare}
\end{figure}

As shown in Fig.~\ref{fig:Imaging_result_compare}, the imaging results obtained by the 4D-FFT algorithm and our proposed algorithm at SNR of -20 dB, -5 dB, and 10 dB are depicted for the qualitative analysis. Horizontal comparisons of Fig.~\ref{fig:Imaging_result_compare} (a)-(c) and (d)-(f) show that the image point cloud density gradually improves with the increase of the SNR, and the imaging effect becomes better gradually; the vertical comparisons show that our proposed ISAC imaging signal processing method based on the 2D-FFT with 2D-MUSIC can obtain higher point cloud density and imaging quality compared with 4D-FFT.

\subsection{Quantitative Analysis of Imaging Results}
\begin{figure}[!htb]
	\centering
	\includegraphics[width=\linewidth]{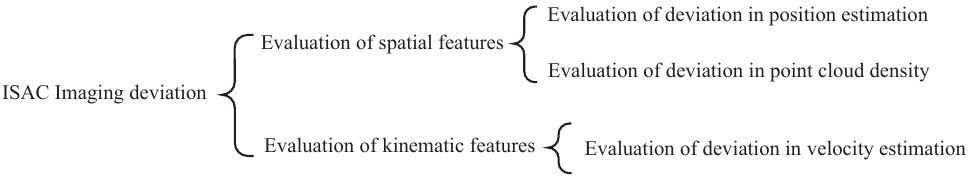}
	\caption{Components of ISAC Imaging Deviation.}
	\label{fig:deviation}
\end{figure}

In order to solve the problem that ISAC 4D imaging quality is difficult to be measured quantitatively, we propose a multi-dimensional evaluation metric for 4D ISAC imaging as shown in Fig.~\ref{fig:deviation}. We measure image quality in terms of overall imaging deviation; the larger the deviation the worse the imaging quality, and vice versa. The overall 4D imaging deviation is divided into two parts: spatial and kinematic features. The spatial deviation is determined by the coordinate difference and the point cloud density difference between the imaging result and the original scene, where we measure the coordinate difference by calculating the Hausdorff distance~\cite{huttenlocher1993comparing} using each of the three 2D projections of the 3D coordinates. The kinematic deviation is measured by Normalized Mean Square Error (NMSE) between the velocity of the main scattering points of the imaging result and the original scene.

\begin{figure}[!htb]
	\centering  %图片全局居中
	\subfigbottomskip=2pt %两行子图之间的行间距
	\subfigcapskip=-5pt %设置子图与子标题之间的距离
	\subfigure[Comparison of various components of 4D ISAC imaging deviation for the two algorithms.]{
		\includegraphics[width=0.45\linewidth]{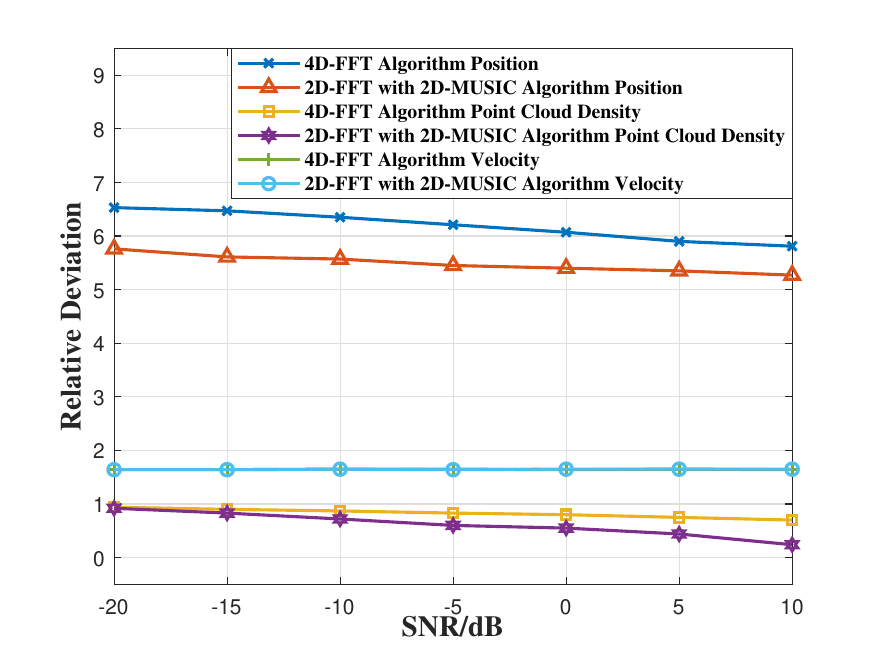}}
	\subfigure[Overall Imaging Deviation Comparison.]{
		\includegraphics[width=0.45\linewidth]{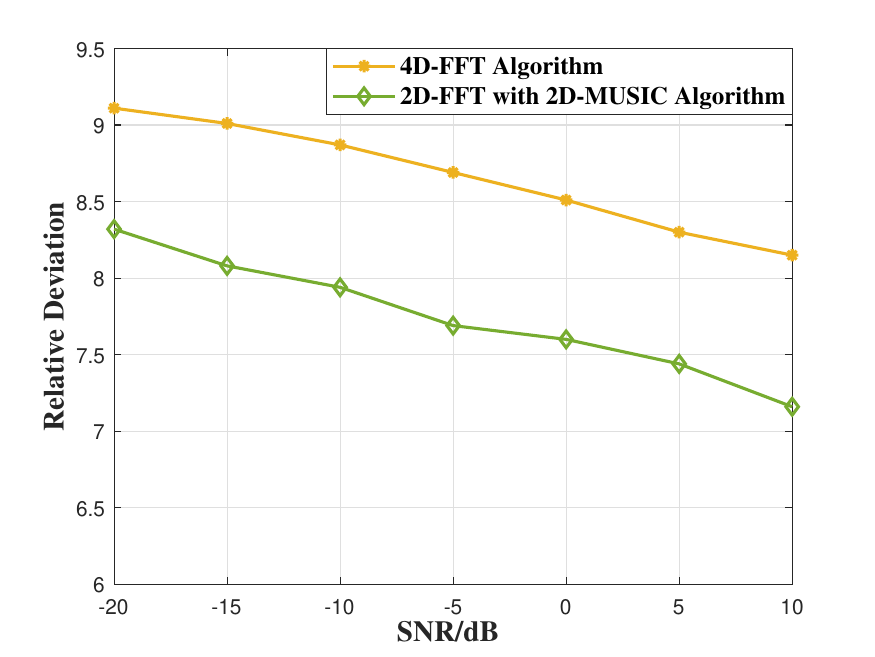}}
	\caption{Variation curve of ISAC imaging deviation with SNR.}
	\label{fig:RD_Compare}
\end{figure}

The trends of the components of the ISAC 4D imaging deviation with SNR are shown in Fig.~\ref{fig:RD_Compare} (a), and the overall trend is shown in Fig.~\ref{fig:RD_Compare} (b). It is worth mentioning that since our proposed algorithm depends on the 2D-FFT algorithm for the estimation of velocity and distance, the trend of kinematic deviation is the same for both algorithms. The results show that the ISAC 4D imaging quality improves with the SNR, and the imaging quality of our proposed algorithm is significantly higher than that of the FFT-based method.

\section{Conclusion}

In this paper, we propose an ISAC 4D imaging system based on 2D-FFT with 2D-MUSIC using 5G DL mmWave signals. We introduce a two-stage CFAR detection algorithm, optimize the DoA estimation and velocity-range estimation processes, and design a transceiver antenna arrangement based on the MIMO virtual aperture technique. Besides,we propose a new evaluation metric for ISAC 4D imaging quality. The simulation results show that our proposed method can obtain ISAC 4D imaging results with higher 4D point cloud density and higher imaging accuracy compared to the traditional methods. 

In the future, we will build ISAC 4D imaging datasets and introduce deep learning models to realize more complex functions.

%\section*{Acknowledgment}

\bibliographystyle{IEEEtran}
\bibliography{references}

\end{document}